\documentclass[12pt, onecolumn]{IEEEtran}
\usepackage{epsfig,epsfig,latexsym,amssymb,amsbsy,cite, graphicx}
\newcommand{\ud}{\mathrm{d}}
\newcommand{\eqn}[1]{Eq.~(\ref{#1})}

\newcommand{\Fig}[1]{Figure~\ref{#1}}

\newcommand{\fig}[1]{Fig.~\ref{#1}}

\begin{document}

\title{Can 100-Gb/s QPSK Signal Locate Adjacent to Legacy 10-Gb/s OOK Signal without Guard-Band?}
\author{Keang-Po~Ho \\
{SiBEAM Inc., Sunnyvale, CA 94085 USA (email: kpho@ieee.org).}}

\maketitle

\begin{abstract}
For 100-Gb/s quadriphase-shift keying (QPSK) signal with on-off keying (OOK) signal in neighboring wavelength-division-multiplexed (WDM) channel, the smoothing filter in the feedforward phase estimation scheme must be optimized to minimize the phase error.
With optimal Wiener filter, typical 0-dBm launched power 10-Gb/s OOK signals give a SNR penalty of 0.66 and 0.30 dB for standard single-mode and nonzero dispersion-shifted fibers, respectively.
\end{abstract}



\section{Introduction}
\PARstart{P}{olarization}-multiplexed (PM) quadriphase-shift keying (QPSK) signals are used for 100-Gb/s long-haul lightwave communication systems \cite{yu10}. 
Typical PM-QPSK signals are used in 50-GHz spacing wavelength-division-multiplexing (WDM) grid, similar to typical legacy 10-Gb/s non-return-to-zero (NRZ) on-off keying (OOK) signals.
An important technical issue is the co-existence of 100-Gb/s PM-QPSK and 10-Gb/s NRZ OOK signals in the same WDM systems.

The effect of adjacent OOK channels to QPSK signal was studied in \cite{tanimura08, bertolini09,  carena09, bononi09, bertran08, kuschnerov08, tao09, piyawanno09}, with many useful results for the issue of cross-phase modulation (XPM) induced nonlinear phase noise.
Simulation was conducted in \cite{bertolini09, carena09, tao09} to find the effect of OOK to QPSK signal. 
As a generic assumption, the simulation may use a random input bit sequence, to find the received constellation and signal-to-noise ratio (SNR) of the QPSK signals. 
The output constellation is rotated to optimize for the SNR, equivalently using phase averaging over the simulation period.
The measurement of \cite{bertran08, tao09} may assume to take the constellation over a period of time that is also rotated to optimize for the SNR, and equivalently using phase averaging over the measurement period.

Phase averaging over a long period of time is not the optimal phase estimation in both simulation and measurement although  the  results are consistent with each other \cite{bertolini09, carena09, bertran08, tao09}.
The study of \cite{bononi09} included phase estimation using a simple 5- or 7-tap averaging filter, far shorter than the equivalent averaging filter of \cite{bertolini09, carena09, bertran08, tao09}.
The group of  \cite{kuschnerov08, piyawanno09} may be the first one to contribute to the idea that carrier phase estimation may reduce the XPM-induced nonlinear phase noise using joint polarization phase estimation.
In all of \cite{kuschnerov08, piyawanno09, bononi09}, phase estimation scheme is not optimized.
Here, for QPSK signals, the optimal filter is designed for the popular feedforward based phase tracking techniques \cite{noe05, ip07, taylor09}.

Wiener filter optimized for laser phase noise was designed for the feedforward phase estimation scheme \cite{ip07, taylor09}.
If the optimal Wiener filter is designed for signal with XPM-induced nonlinear phase noise from adjacent legacy NRZ OOK signals, for SNR penalty of 1 dB, the launched power of NRZ OOK channels can increase over 10 dB compared with the case with the usage of long averaging filter. 

\section{XPM-Induced Nonlinear Phase Noise}

Pump-probe model is typical used to study XPM-induced nonlinear phase noise between two WDM channels \cite{ho0404, tao09, chiang96, hui99, bononi09}.
The power spectral density of XPM-induced nonlinear phase noise from channel 2 (pump) to channel 1 (probe) is
\begin{equation}
\Phi_{\phi_1}(f) = \Phi_{P_2}(f) |H_{12}(f)|^2,
\end{equation}
where $\Phi_{P_2}(f)$ is the power spectral density of the intensity of the pump, and $H_{12}(f)$ is the transfer function.
With OOK signal,  the spectral density of $\Phi_{P_2}(f)$ is 
$\Phi_{P_2}(f) = P_0^2 T_b \mathrm{sinc}^2 (f T_b),$ where $T_b$ is bit interval

With one fiber span, the transfer function is \cite{ho0404, tao09}
\begin{equation}
H_{12}(f) = 2 \gamma \frac{1 - e^{-\alpha L + j 2 \pi f d_{12} L}}{\alpha -j 2 \pi f d_{12}}.
\label{XPMH12}
\end{equation}
where $\gamma$ is the fiber nonlinear coefficient, $\alpha$ is fiber attenuation coefficient, $L$ is the fiber length, $d_{12} \approx D \Delta \lambda$ is the relative walk-off between two channels with wavelength separation of $\Delta \lambda$, and $D$ is the dispersion coefficient.
The transfer function of \eqn{XPMH12} ignores the distortion of the pump in the fiber \cite{hui99, bononi09}.
As the pump is only 10-Gb/s channel and channel spacing is 50 GHz, channel walk-off has far larger effect than pump distortion \cite{tao09}.

With $K$ fiber spans, the transfer function becomes \cite{ho0404}
\begin{equation}
H_{12}(f) = 2 \gamma \frac{1 - e^{-\alpha L + j 2 \pi f d_{12} L}}{\alpha -j 2 \pi f d_{12}} \times \frac{1 - e^{ - j 2 \pi f (1 - \kappa) d_{12} K L} } {1 - e^{ - j 2 \pi f (1 - \kappa) d_{12} L}}.
\label{XPMH12K}
\end{equation}
where $\kappa$ is the fraction of optical dispersion compensation per span, i.e., $\kappa = 1$ and $\kappa = 0$ for perfect and without optical dispersion compensation, respectively \cite{ho0404}. 
The transfer function of \eqn{XPMH12K} assumes $K$ cascaded identical fiber spans with the same configuration without loss of generality.

%
%

\section{Feedforward Phase Estimation}

\begin{figure}[t]
\center{
\includegraphics[width=.8 \textwidth]{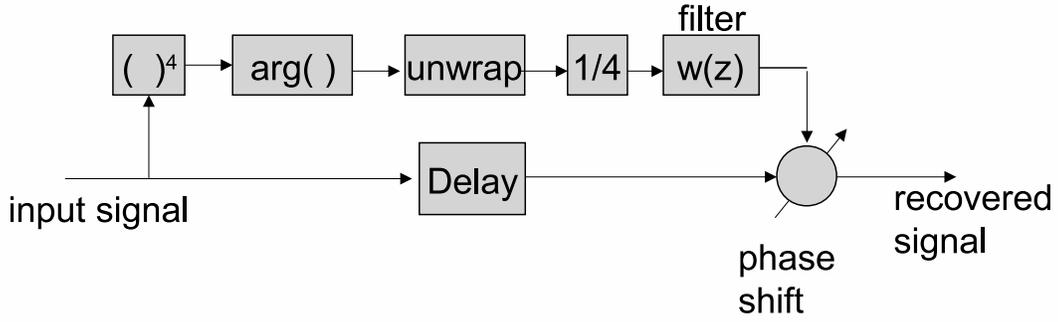}}
\caption{Feedforward carrier recovery for QPSK signals.}
\label{fig:ff}
\end{figure}

Feedforward phase estimation \cite{noe05, ip07, taylor09} is typically used for high-speed QPSK signals.
\Fig{fig:ff} shows the schematic diagram of feedforward carrier recovery for QPSK signals.
The signal is first raised to the 4th power to obtain the phase without modulation,  unwrap the phase, taking the factor of $1/4$, and smoothing using a filter of $W(f)$, to compensate for the phase variations.
The optimal smoothing filter of $W(f)$ is designed here for system with XPM-induced nonlinear phase noise.
The filter $W(f)$ is expressed as $w(z)$ in \fig{fig:ff} to emphasize the discrete time operation of the filter.
Because the transfer function of \eqn{XPMH12K} is a low-pass response, there is almost no numerical difference between continuous- and discrete-time analysis of the system.
Continuous-time analysis is used here.

If the received signal is denoted as $A e^{j \theta_r + j \phi_e + j \theta_n}$ where $A$ is the amplitude,  $\theta_r = (2k+1) \pi/4$ with $k = 0, \dots, 3$ as the transmitted phase, $\phi_e$ is the phase noise, and $\theta_n$ is the phase due to additive Gaussian noise.
The phase of $\theta_n$ is independent of the phase noise $\phi_e$.
The input to the smoothing filter $W(f)$ is $\phi_e + \theta_n$.
The variance of $\theta_n$ is $\sigma^2_{\theta_n} = 1/2\rho_s$ when $\rho_s$ is larger than 10 dB \cite[Fig. 4.A.1]{hobook}.
The output of the smoothing filter should be $\hat{\phi}$ as an estimation of $\phi_e$.
From the theory of Wiener filter for smoothing \cite[Sec. 13-3]{papoulis2}, the optimal smoothing filter is
\begin{equation}
W(f) = \frac{\Phi_{\phi_e}(f)}{\Phi_{\phi_e}(f) + N_{\theta_n}},
\label{wfwiener}
\end{equation}
where $\Phi_{\phi_e}(f)$ is spectral density of the phase noise, and $ N_{\theta_n}$ is the spectral density of $\theta_n$.
Although the smoothing filter \eqn{wfwiener} is non-casual, the delay in the main signal path may be used to transfer $W(f)$ to casual filter \cite{taylor09}.

The phase estimation  mean-square error (MSE) is $\mathcal{E} = E\{ (\hat{\phi} - \phi_e)^2 \}$ or
\begin{equation}
\mathcal{E} =  \int \left| 1 - W(f) \right|^2 \Phi_{\phi_e}(f) \ud f +  N_{\theta_n}\int \left|W(f) \right|^2 \ud f,
\label{ergen}
\end{equation}
where the integration intervals here or later are all from $-\infty$ to $+\infty$. 
With the smoothing filter of \eqn{wfwiener}, we obtain
\begin{equation}
\mathcal{E}_{\mathrm{min}} = \int \frac{ \Phi_{\phi_e}(f) N_{\theta_n}}   { \Phi_{\phi_e}(f) +  N_{\theta_n}} \ud f.
\label{erwiener}
\end{equation}
With a very long averaging window, the second term of \eqn{ergen} is equal to zero and the phase error MSE is approximately equal to  $ \int \Phi_{\phi_e}(f) \ud f$.

\section{Numerical Results}

The effects of XPM-induced nonlinear phase noise are studied for the middle channel of a 50-GHz spacing 81-channel WDM systems with lower-band 41 QPSK channels and upper-band 40 NRZ OOK channels.
Required for the 10 Gb/s NRZ OOK signals, the optical dispersion compensation per span is $\kappa = 1.05$ and $\kappa = 0.78$ for single-mode fiber (SMF) with $D = 17$ ps/km/nm and non-zero dispersion-shifted fiber (NZDSF) with $D = 3.8$ ps/km/nm, respectively, similar to that in \cite{carena09}.
The QPSK signal has two polarizations each with a symbol rate of 28 GHz, providing an overall data rate of 100 Gb/s after error correction.
The WDM system has 20 spans with $L = 90$ km per span, loss coefficient $\alpha =$ 0.2 dB/km, nonlinear coefficient of $\gamma = 1.3$ and $1.5$ /W/km for SMF and NZDSF, respectively.

\begin{figure}[t]
\center{
\includegraphics[width=.6 \textwidth]{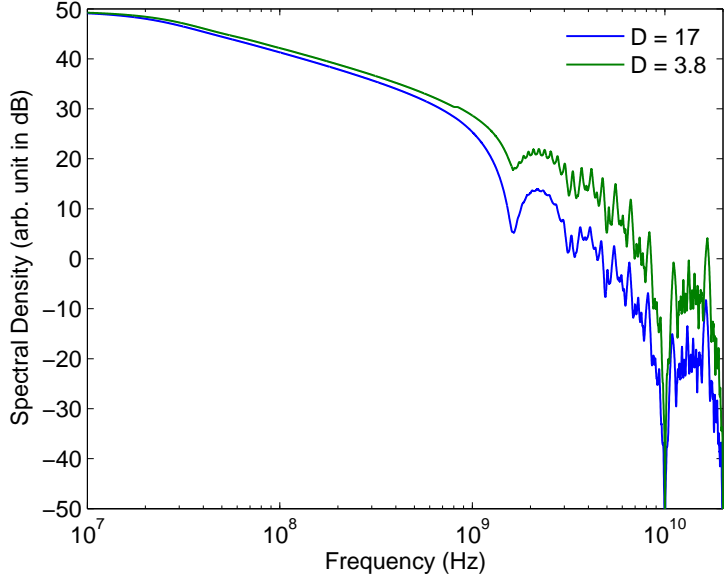}}
\caption{The spectral density  $\Phi_{\phi_e}(f)$ of XPM-induced nonlinear phase from NRZ OOK channels. 
}
\label{fig:spnrz}      
\end{figure}

\Fig{fig:spnrz} shows the spectral density of the phase error $\Phi_{\phi_e}(f)$ due to XPM-induced nonlinear phase noise from all 40 NRZ OOK channels to the first QPSK channel.
Mostly in the frequency less than 1 GHz, the phase error can be effectively reduced by a Wiener filter.
In the frequency less than 1 GHz, $W(f)$ is approximately equal to 1 from \eqn{wfwiener} that fully eliminate the phase noise by the factor of $|1 - W(f)|^2$  at low frequency.
 
\begin{figure}[t]
\center{
\includegraphics[width=0.6 \textwidth]{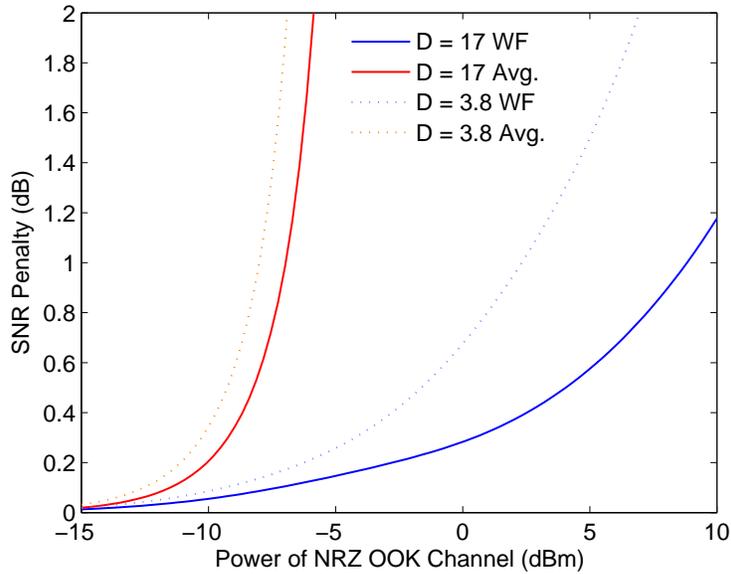}}
\caption{SNR penalty due to XPM-induced nonlinear phase noise as a function of NRZ OOK channel power.
}
\label{fig:qpsknrz}      
\end{figure}

\Fig{fig:qpsknrz} shows the SNR penalty as a function of the power of the NRZ OOK signal. 
The SNR penalty is calculated by expanding the phase distribution of Gaussian noise as a Fourier series \cite[App. 4.A]{hobook}, similar to both \cite{ho0404, bononi09}.
With an independently optimized launched power, the equivalent SNR of the QPSK channel is assumed to be 10 dB, for an error probability of about $10^{-3}$.
The 10-dB SNR should include all the contributions from amplifier noise, the nonlinearities by itself and other QPSK channels.

Using long-time averaging for phase tracking, the launched power of NRZ OOK channels must be less than -7 and -8 dBm for SMF and NZDSF, respectively, for a SNR penalty less than 1 dB.
If the optimal Wiener filter is used, the launched power of NRZ OOK channel may be increased to over 8 and 2 dBm for SMF and NZDSF, respectively, representing an improvement of 15 and 10 dB.
Typical legacy NRZ OOK channel has launched power about 0 dBm per channel, giving a penalty of only about 0.66 and 0.30 dB for SMF and NZDSF, respectively.

\Fig{fig:qpsknrz} assumes that the NRZ OOK signals are in only one-side of the QPSK signal without guard-band.
For the case that a QPSK signal is in the middle of NRZ OOK signals, the XPM-induced nonlinear phase noise is doubled in the worst case, equivalently the curves of \fig{fig:qpsknrz} are shifted by 1.5 dB in the $x$-axis.
For PM-QPSK signal, the results of \fig{fig:qpsknrz} assume the NRZ OOK channel is aligned with the worst polarization sub-channel.
On average, the curves of \fig{fig:qpsknrz} may be shifted by $-0.4$ dB. 

\section{Conclusion}

Using Wiener filter to smooth the XPM-induced nonlinear phase noise, NRZ OOK channel can be located adjacent to PM-QPSK channel without guard-band.
Compared using a long-time averaging for phase estimation, the usage of optimal Wiener filter can increase the NRZ OOK launched power by over 15 and 10 dB for SMF and NZDSF, respectively.



\end{document}